# A Cost-based Storage Format Selector for Materialization in Big Data Frameworks

Rana Faisal Munir · Alberto Abelló ·
Oscar Romero · Maik Thiele · Wolfgang
Lehner

**Abstract** Modern big data frameworks (such as Hadoop and Spark) allow multiple users to do large-scale analysis simultaneously. Typically, users deploy Data-Intensive Workflows (DIWs) for their analytical tasks. These DIWs of different users share many common parts (i.e, 50-80%), which can be materialized to reuse them in future executions. The materialization improves the overall processing time of DIWs and also saves computational resources. Current solutions for materialization store data on Distributed File Systems (DFS) by using a fixed data format. However, a fixed choice might not be the optimal one for every situation. For example, it is well-known that different data fragmentation strategies (i.e., horizontal, vertical or hybrid) behave better or worse according to the access patterns of the subsequent operations.

In this paper, we present a cost-based approach which helps deciding the most appropriate storage format in every situation. A generic cost-based storage format selector framework considering the three fragmentation strategies is presented. Then, we use our framework to instantiate cost models for specific Hadoop data formats (namely SequenceFile, Avro and Parquet), and test it with realistic use cases. Our solution gives on average 33% speedup over SequenceFile, 11% speedup over Avro, 32% speedup over Parquet, and overall, it provides upto 25% performance gain.

Rana Faisal Munir
E-mail: fmunir@essi.upc.edu

Alberto Abelló
E-mail: aabello@essi.upc.edu

Oscar Romero
E-mail: oromero@essi.upc.edu

Maik Thiele
E-mail: maik.thiele@tu-dresden.de

Wolfgang Lehner
E-mail: wolfgang.lehner@tu-dresden.de

# 1 Introduction

Data analysis plays a decisive role in today's data-oriented organizations, which increasingly produce and store larger volumes of data in the order of petabytes to zettabytes [24]. Storing and processing such large volumes of data heavily rely on the use of distributed frameworks. Within the last few years, many frameworks such as Apache Hadoop[1] and Apache Spark[2] appeared to process large data volumes and deploy complex analytical workflows that orchestrate multiple tasks, where each produces an output that is used as input for subsequents. These anlytical workflows have many redundant tasks, whose materialization can bring the benefits of re-usage and improve the execution time.

An in-depth study of seven enterprises [6] shows that 80% of different analytical workflows have common tasks, and materializing their output would clearly give performance gains in future executions. Thus, the study shows the importance of materializing the output of repetitive tasks. However, it also raises two questions: "*(1) which tasks should be materialized?*" and "*(2) what type of layout should be used when persisting their output?*".

Answers to the first question have already been given. In [8,14,15,19,21,25], authors provide tools which are choosing the common tasks to be materialized so that the overall analytical workflows speed up. Typically, the output of chosen tasks to be materialized are directly stored on *Distributed File System (DFS)*. Unfortunately, the existing materialization solutions use a single fixed layout and completely ignore the second question "*what type of layouts should be used when persisting their output?*". Furthermore, DFS I/O operations are expensive and the load time can be reduced if the physical layout is chosen based on their subsequent use. Obviously, a fixed storage layout can not be optimal for all types of workloads. Indeed, [1] shows the importance of storing data according to their access pattern and that single fixed layouts are not good for all types of workloads. Similarly, [1,11,16] also focus on the importance of storing data according to their access patterns and highlight the effect of different storage layouts on different workloads[3]. Nevertheless, still no current solution lets us choose the layout in an automatic fashion.

In this paper, we present a cost-based approach to address the second question and find the most appropriate storage layout for materializing the output of common tasks. However, as a cost model requires statistical information about the data and their analytical flows in order to make a decision, we also consider using a rule-based one for cold-starts. Accordingly, we first apply rules for choosing storage layouts, while collecting the statistical information.

---

[1] http://hadoop.apache.org  
[2] https://spark.apache.org  
[3] http://www.svds.com/how-to-choose-a-data-format

Once the required statistical information has been gathered, we could apply the cost-model.

Our contributions are as follows:

- We present a generic I/O cost model for the three fragmentation strategies (i.e., horizontal, vertical, and hybrid) in big data frameworks, for estimating their read and write costs.
- We instantiate the cost model on *Hadoop Distributed File System (HDFS)*, for SequenceFile, Avro, and Parquet.
- We propose and implement a generic framework for big data frameworks, to materialize the selected results in the appropriate data format.
- We conduct experiments on two de-facto standard industry benchmark for Decision Support System (DSS) to test our approach. Our results show that we effectively manage to reduce the load time of materialized results compared to any single fixed layout, by providing upto 25% average performance gain.

The remainder of this paper is organized as follows: In Section 2, we discuss the storage layouts and our motivation. In Sections 3 and 4, we discuss our approach and the generic cost model in detail. In Section 5, we show our experimental results. In Section 6, we discuss the related work. Finally, in Section 7, we conclude the paper.

## 2 Background and Motivation

In this section, we discuss the different storage layouts available and their corresponding instantiation for HDFS. Moreover, we motivate our work by illustrating the fixed layout limitations.

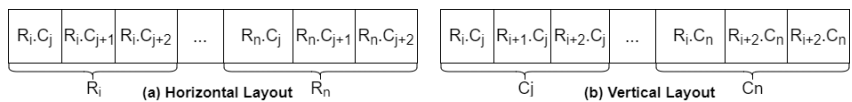

**Fig. 1** Horizontal and vertical layouts

### 2.1 Storage layouts

There are many layouts, used in different processing frameworks, that can be divided into three categories based on how they fragment data: horizontal, vertical or hybrid. Each concrete layout has its own physical storage structure that is beneficial for a specific kind of workloads.

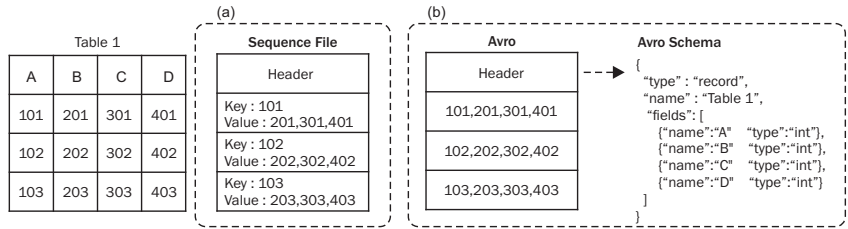

**Fig. 2** Examples of SequenceFile and Avro layouts

*Horizontal layouts* They are organized row-wise, and the data of each row are stored together, as shown in Figure 1a (where $R$ represents the row and $C$ represents the column of a row). For this reason, a horizontal layout especially suits scan-based workloads. However, if a query is just referring to a small subset of columns, this layout results in a low usefulness rate of data read, since non-required columns will be fetched anyway. In HDFS, the horizontal layout is implemented by SequenceFile[4] and Avro[5]. SequenceFile is a special type of horizontal layout storing simple key-value data, whereas Avro explicitly splits data into columns inside every row. Figure 2 shows an example of a table and its corresponding format in SequenceFile (i.e., Figure 2a) and Avro (i.e., Figure 2b).

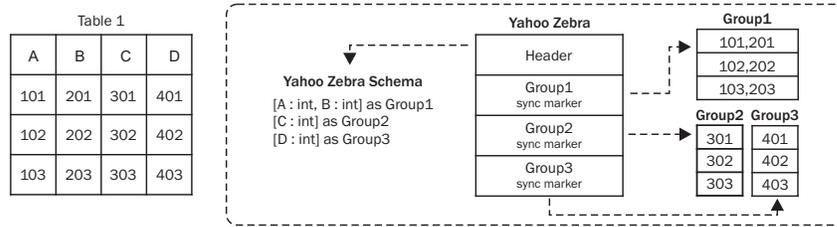

**Fig. 3** Example of Zebra layout

*Vertical layouts* They divide each row into columns, and store each column separately, which is beneficial for workloads reading just a few columns. Thus, these layouts excel in projection-based workloads. Figure 1b sketches the physical structure of vertical layouts. Zebra[6], illustrated in Figure 3, is an implementation of this kind for HDFS.

*Hybrid layouts* They are a combination of horizontal and vertical layouts, having two alternative implementations: Either the data is divided horizontally and then vertically, like in Figure 4a, or vice versa, like in Figure 4b. Both cases are especially helpful for combinations of projection and selection operations. There are many implementations of this kind, but the most popular ones in HDFS are Optimized Row Columnar (ORC)[7] and Parquet[8], both primarily fragmenting data horizontally. Figure 5 exemplifies Parquet.

---
[4] https://wiki.apache.org/hadoop/SequenceFile
[5] https://avro.apache.org
[6] https://wiki.apache.org/pig/zebra

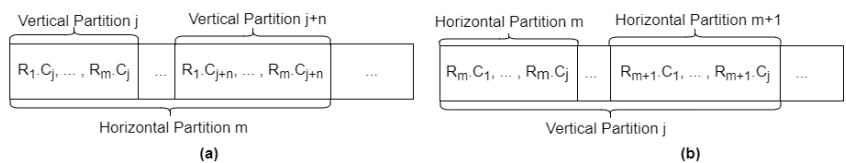

**Fig. 4** Hybrid layouts

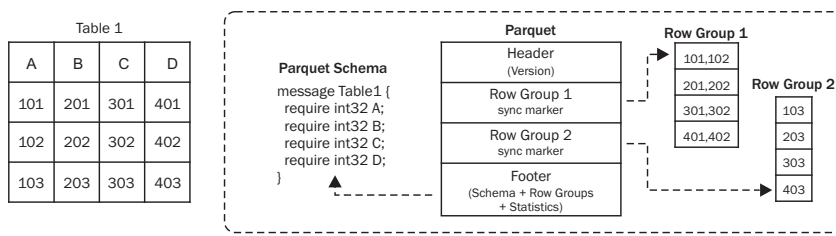

**Fig. 5** Example of Parquet layout

2.2 Existing materialized solutions

As discussed previously, there are many available materialization solutions [8, 14, 19, 21, 25] for big data frameworks, which can be used to select common parts for materialization. In this paper, we use ReStore [8] because it is a simple but powerful solution based on heuristic rules. Importantly, there is an available implementation[9] of ReStore in Apache Pig[10], which we used for our experiments. However, our approach is not tied to any materialization solution and ReStore could be replaced by other more sophisticated methods if required. The heuristics of ReStore are categorized into conservative and aggressive: *Conservative heuristics* aim at materializing the outputs of those operators, i.e., Projection and Selection, which reduce the size of the data. Whereas, *aggressive heuristics* materialize the outputs of those operators, i.e.,

---

[7] https://orc.apache.org
[8] http://parquet.apache.org
[9] https://github.com/ami07/ReStoreV2
[10] https://pig.apache.org

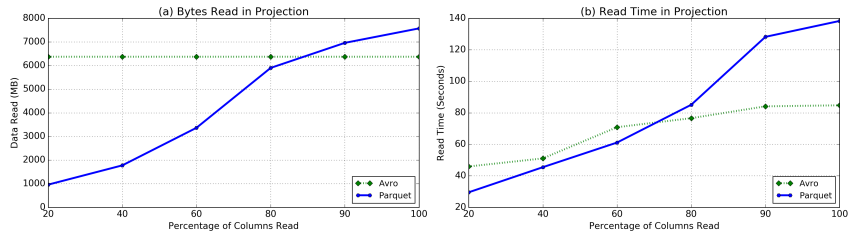

**Fig. 6** The effect of the number of retrieved columns on different layouts

Join, Group By, which are computation-intensive. Section 5 discusses the results of ReStore for TPC-H[11] and TPC-DS[12].

2.3 Layout performance comparison

Ad-hoc and exploratory analysis are very popular among data analysts, helping them to understand different aspects of their business. However, it is very difficult to tune a system for such scenarios since the workload is very dynamic, and *current solutions are not considering layouts depending on frequency of operations, and ignore this fact when materializing the output of redundant tasks.*

To illustrate the drawback of the current static approaches let us assume the following example from TPC-H. Lets assume the result of the join between *Lineitem* and *Part* tables is chosen to be a materialized. Figure 6 shows the response time for such materialized result for horizontal and hybrid layouts using a simple projection-based query. It can be seen that Parquet (i.e., a hybrid layout) performs well when the total amount of data read from disk is below 75%, whereas Avro (i.e., a horizontal layout) performs better as soon as we read more than 75% of data. Thus, this shows that the characteristics of the query help to determine the optimal layout.

## 3 Our Approach in a Nutshell

From here on, a *Data-Intensive Workflow (DIW)* is represented as a directed-acyclic graph of operations (an example can be seen in Figure 11). Nodes represent operations and directed edges show the dependencies between the nodes. The starting node of an edge produces the data to be consumed by the ending node (note that a node output can be consumed by several nodes). Different DIWs can have multiple common nodes, whose output is referred to as *Intermediate Results (IR)*.

Given a DIW, Figure 7 illustrates the flowchart of our approach. Following the two questions introduced in Section 1, first, (i) it decides which IR to materialize using ReStore and, then, (ii) for each of them which storage layout to choose.

---
[11] http://www.tpc.org/tpch
[12] http://www.tpc.org/tpcds

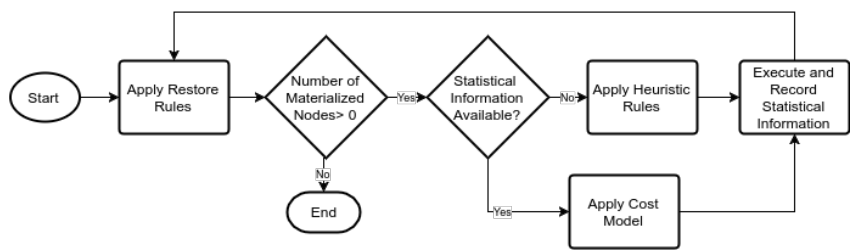

**Fig. 7** Flowchart of our approach

3.1 Storage layout selection

Since existing materialized solutions apply a fixed layout format for materialized nodes, our approach helps them to decide the best storage layout for each node chosen to be materialized. If statistical information about the node is available, we use the cost-based model to decide the storage layout. If for any reason we have not enough statistical information (see the variables required by our cost model in Section 4) to make a decision on a given node, we can still apply heuristic rules [20] to determine the storage format. The heuristic rules choose a storage layout based only on the operation type. Obviously, it might happen that the heuristic rules do not choose the best storage layouts since they do not consider essential information to estimate the total volume of data to be read from the disk. For example, projections/selections can perform differently based on the percentage of columns/rows read. Thus, factors such as the number of columns and selectivity factors may drastically impact on the operation performance depending on the storage layouts, as illustrated in Figure 6. These factors cannot be considered in heuristic rules, because heavily depend on the concrete operation and data characteristics. Still, heuristic rules provide a fair first-approach to the problem with small computational requirements in scenarios where there is lack of information. Oppositely, if the required statistics are available, the cost-based approach, like the one in Section 4, is more accurate. Finally, the DIW is executed and materializes the chosen nodes with their chosen storage formats. Also, it records/updates the needed statistical information to be used in the future.

## 4 Cost-Based Model

The cost-based model relies on a wide range of statistical information that is summarized in Table 1, containing system constants, data statistics, workload statistics as well as layout variables. We assume the constants depending on the configuration of the environment (e.g., $BW_{disk}$, $BW_{net}$) are given and the statistics are collected during the DIW execution. Moreover, it should be noted that we consider only I/O cost in our cost model, because it is the dominant factor in DIWs.

---
[13]Extra 4 bytes are considered for variable length columns

Table 1 Parameters of the Cost Model

| Variable | Description |
|---|---|
| *System Constants* | |
| $R$ | Replication factor |
| $p$ | Probability of accessed replica being local |
| $Size(Chunk)$ | Block size in the DFS |
| $BW_{disk}$ | Disk bandwidth |
| $BW_{net}$ | Network bandwidth |
| $Time_{seek}$ | Disk seek time |
| $Time_{disk}$ | $\dfrac{Size(Chunk)}{BW_{disk}}$ |
| $Time_{net}$ | $\dfrac{Size(Chunk)}{BW_{net}}$ |
| *Data Statistics* | |
| $|IR|$ | Number of Rows in IR |
| $Size(Row)$ | Average Row Size of IR |
| $Size(Col)^{13}$ | Average Column Size of IR |
| $Cols(IR)$ | Columns of IR |
| *Workload Statistics* | |
| $RefCols(IR)$ | Number of columns used in an operation |
| $SF$ | Selectivity factor of an operation |
| *Layout Variables* | |
| $Size(RowGroup)$ | Row group size of hybrid layouts |
| $Size(Meta_{layout})$ | Meta data size for a given layout |
| $Size(Body)$ | Size of the body of the file |
| $Size(Header)$ | Size of the header of the file |
| $Size(Footer)$ | Size of the footer of the file |

$$Size(Layout) = Size(Header_{layout}) \qquad (1)$$
$$+ Size(Body_{layout})$$
$$+ Size(Footer_{layout})$$

$$Used_{chunks}(Layout) = \frac{Size(Layout)}{Size(Chunk)} \qquad (2)$$

$$Seeks(Layout) = \lceil Used_{chunks}(Layout) \rceil \qquad (3)$$

Independently of the kind of layout, the driving factor of our cost model is the file size. The body, together with the header and footer compose it (Equation 1). From that, we can obtain the number of chunks used (Equation 2) and the number of disk seeks we need to reach them (Equation 3). The number of seeks is equal to the total number of chunks rounded up, because one seek is required for every chunk, even if it is not full. Note that modern Solid State Disks (SSDs) also have seek time (i.e., time required to turn on the right circuit), however their seek time is much less (i.e., 0.1ms) compared to hard disks (i.e., 0.8ms)[14].

In the next subsections, we analyze the cost of data writes and reads, because they are the dominant factors in the overall execution time of operations. The write cost model estimates the data volume footprint of each layout as well as the cost incurred in writing it, while the read cost model estimates the cost of an operation depending on the access pattern. Regarding the latter, given the simplicity of a file system (far from that of a DBMS) only three operations are possible (namely full scan, projection, and selection).

$$W_{WriteTransfer} = \frac{Time_{disk} + (R-1)*Time_{net}}{Time_{seek} + Time_{disk} + (R-1)*Time_{net}} \quad (4)$$

$$Cost_{write}(Layout) = Used_{chunks}(Layout) * W_{WriteTransfer} \\ + Seeks(Layout) * (1 - W_{WriteTransfer}) \quad (5)$$

4.1 Write cost

First of all, we have to take into consideration that distributed processing frameworks are using DFS to store data into multiple chunks. Thus, the number of chunks of a file is used to estimate the overall writing costs. Given that a chunk consists of multiple contiguous disk blocks and inside it, sequential read is guaranteed, assuming that the chunk size is smaller than a disk cylinder, the write cost can be simply computed as the number of chunks plus the seek cost to locate the position of each. Nevertheless, since our cost model is thought for distributed processing frameworks, we further need to consider the replication factor $R$ used for fault-tolerance, and therefore the network costs for writing $R$ copies needs to be taken into account. We assume that the replication procedure is sequential (as it is in HDFS) and the multiple copies are written one after another. Equation 4 gives the weight of transferring a chunk by considering the network and the disk write against the seek costs. Finally, Equation 5 shows the total write cost taking both seek and transfer weights into account.

In the following, we present the write cost for each horizontal, vertical and hybrid layouts.

*Horizontal layouts* They store data row-wise into the body section. Oppositely, metadata containing information such as schema and version, is written into the header and footer sections. Nevertheless, in some implementations, additional metadata is also written in the body with every row, for example, metadata used to separate each row or each column (i.e., its size is not constant and depends on the number of columns).

---

[14] http://www.ieee802.org/3/CU4HDDSG/public/sep15/Kipp_CU4HDDsg_01a_0915.pdf

$$Size(Body_{horizontal}) = (Size(Row) + Size(Meta_{HRow})) * |IR| \quad (6)$$
$$+ Size(Meta_{HBody})$$

Equation 6 estimates the size of the body by multiplying the average row size and metadata by the total number of rows, plus other metadata we may find in the body section.

$$Size(OneColWithMeta) = Size(Col) * |IR| + Size(Meta_{VBody}) \quad (7)$$

$$Size(Body_{vertical}) \quad = Size(OneColWithMeta) * Cols(IR) \quad (8)$$

*Vertical layouts* They store each column independently (i.e., values of a column, which share the same data type, are stored consecutively) using a separator of fixed size between columns. Equation 7 provides the estimation of the individual column size, which is used in Equation 8 to determine the overall size of the body by multiplying the size of one column by the total number of columns.

$$Used_{RG}(Hybrid) = \frac{(Size(Col) * |IR| + Size(Meta_{YCol})) * Cols(IR)}{Size(RowGroup)} \quad (9)$$

$$Size(Meta_{hybrid}) = \lceil Used_{RG}(Hybrid) \rceil * Size(Meta_{YRowGroup}) \quad (10)$$

$$Size(Body_{hybrid}) = Used_{RG}(Hybrid) * Size(RowGroup) \quad (11)$$
$$+ Size(Meta_{hybrid})$$

*Hybrid layouts* They are a combination of horizontal and vertical layouts. They divide rows into horizontal partitions known as row groups and each row in one row group is further divided into vertical partitions storing each column separately, and inserting metadata between them. Additionally, they also store metadata for every row group. Thus, the total size of the body depends on the number of row groups being used, which can be estimated as in Equation 9 and the size of metadata of row groups is estimated in Equation 10. Notice that the metadata of the row group is stored irrespectively of it being completely full, so this must be rounded up. Furthermore, Equation 11 obtains the size of the body by multiplying the number of row groups by the size of a row group and by adding the total size of metadata.

4.2 Read cost

This section presents the read cost model for scan, projection and selection operations. All DIW operations in current massively distributed processing environments use a full scan access pattern on the DFS, except projection and selection operations that are specifically supported natively in some storage layouts. Thus, we consider them separately in the following.

$$Size(Scan_{layout}) = Size(Layout) \\ + (Used_{chunks}(Layout) * Size(Meta_{layout})) \quad (12)$$

**Scan** reads all stored data from the disk, irrespective of the layout being used. Additionally, the metadata (such as schema, statistics, etc.) stored inside header or footer sections, reads separately in each task. The reason is that the distributed processing engines (such as Hadoop and Spark) create a separate process for each task with its own memory. This memory is not accessible to other tasks and hence, forces to read all metadata in each task separately, and consequently, increases the reading size. The number of tasks is equal to the number of used chunks. Equation 12 estimates the scan size, which can be used further to estimate the scan cost.

The scan cost purely depends on the number of used chunks to be read. Assuming the block is the transfer unit between disk and memory, there are three factors impacting the cost: the average seek time needed to locate a disk block cylinder, the rotation time to move the disk head over the cylinder to reach the block, and the transfer time to bring data in the block from disk into memory. Nevertheless, despite every chunk consists of multiple blocks on disk, it should be noted that DFS typically guarantee that all disk blocks are contiguous within one disk cylinder, under the assumption that the chunk size does not go beyond the cylinder size. This is why we do not need to consider seek time for all the disk blocks. Instead, we only consider seek time once for every chunk. Also, as confirmed in our experiments, the rotation time is negligible, because modern hardware and operating systems implement very effective pre-fetching techniques. Furthermore, our cost model is also applicable to SSDs. Since SSDs have very small seek time and high I/O speed, the corresponding system constants need to be replaced. For the rest, since the basic unit of our cost model is defined in terms of bytes, all the estimations will remain the same.

$$W_{ReadTransfer} = \frac{Time_{disk} + (1-p) * Time_{net}}{Time_{seek} + Time_{disk} + (1-p) * Time_{net}} \quad (13)$$

$$Used_{chunks}(Scan_{layout}) = \frac{Size(Scan_{layout})}{Size(Chunk)} \quad (14)$$

$$Cost_{scan}(Layout) = Used_{chunks}(Scan_{layout}) * W_{ReadTransfer} \\ + Seeks(Layout) * (1 - W_{ReadTransfer}) \quad (15)$$

On the other hand, we have to take under consideration that in a distributed data processing framework data can be accessed remotely. Consequently, we introduce a probability $p$ to indicate the likelihood of chunks being accessed locally (i.e., data shipping through the network is not needed to reach the operation executor). This is used to estimate the weight of transferring the chunk data compared to the corresponding seek time using Equation 13. Then, Equation 14 estimated the total number of read chunks and Equation 15 provides the scan cost taking both the seek and the transfer cost into account with the corresponding weights.

**Projection** helps in fetching only some columns from disk (skipping others) to save some I/Os. Its cost depends on the support provided by each layout.

*Horizontal layouts* They do not provide specific support for projection operation, but actually use a full scan to bring all the data into memory and only afterwards discard the unnecessary columns. Therefore, its cost is exactly the same as that of scan (i.e., Equation 15).

$$Size(Project_{vertical}) = Size(Header_{vertical}) + Size(Footer_{vertical}) \quad (16)$$
$$+ Size(OneColWithMeta) * RefCols(IR)$$

$$Cost_{project}(Vertical) = Used_{chunks}(Project_{vertical}) * W_{ReadTransfer} \quad (17)$$
$$+ RefCols(IR) * Seeks(OneColWithMeta)$$
$$* (1 - W_{ReadTransfer})$$

*Vertical layouts* They do support projections. Their cost depends on the size retrieved data, which is exactly that of the referred columns and the metadata in the header and footer sections, as in Equation 16. The seek time depends on the number of retrieved columns (that might not be consecutively stored in disk), and their size. Equation 17 combines both components considering the weight of a read transfer as defined in Equation 13.

$$Used_{rows}(RowGroup) = \frac{|IR|}{Used_{RG}(Hybrid)} \tag{18}$$

$$\begin{aligned}Size(RefCols) &= (Size(Col) * Used_{rows}(RowGroup) \\ &\quad + Size(Meta_{YCol})) * RefCols(IR)\end{aligned} \tag{19}$$

$$\begin{aligned}Size(Project_{hybrid}) &= Size(Header_{hybrid}) + Size(Footer_{hybrid}) \\ &\quad + (Size(RefCols) + Size(Meta_{YRG})) \\ &\quad * Used_{RG}(Hybrid) \\ &\quad + (Used_{chunks}(Hybrid) * Size(Meta_{hybrid}))\end{aligned} \tag{20}$$

$$\begin{aligned}Cost_{project}(Hybrid) &= Used_{chunks}(Project_{hybrid}) * W_{ReadTransfer} \\ &\quad + Seek(Hybrid) * (1 - W_{ReadTransfer})\end{aligned} \tag{21}$$

*Hybrid layouts* They also natively support projection, and similarly to vertical layouts, we have to calculate its size to estimate the cost. However, hybrid layouts store data into multiple row groups. Therefore, we first need the row group size to estimate the projection size. As each row group contains a subset of rows, we estimate it as in Equation 18. Furthermore, Equation 19 gives the size of referred columns in a group, which is then used in Equation 20 to estimate the overall projection size. Similar to the scan cost, hybrid layout also reads metadata separately for projection in each task, which we consider in the projection size. Hybrid layouts also have a seek cost to be considered, which depends on the number of row groups needed by the overall size of the file (not only of the result of the projection). Similar to previous cases, we can estimate the projection cost of hybrid layouts by appropriately weighting the transfer and seek times as in Equation 21.

**Selection** helps in fetching only some rows from disk (skipping others) to save some I/Os. As for projection, its cost depends on the support provided by each layout.

*Horizontal and vertical layouts* They do not natively support this operation. They perform scan to bring all the data into memory and then filter them out based on the given predicate. Thus, their selection cost is the same as that of scan.

$$P(RGSelected) = 1 - (1 - SF)^{Used_{rows}(RowGroup)} \quad (22)$$

$$Size(RowsSelected) = (Size(Col) * SF * |IR| + Size(Meta_{YCol})) \quad (23)$$
$$* Cols(IR)$$

$$Used_{RG}(Select_{hybrid}) = \begin{cases} Used_{RG}(Hybrid) \\ *P(RGSelected) & \text{if Unsorted} \\ \\ \left\lceil \dfrac{Size(RowsSelected)}{Size(RowGroup)} \right\rceil & \text{if Sorted} \end{cases} \quad (24)$$

$$Size(Select_{hybrid}) = Size(Header_{hybrid}) + Size(Footer_{hybrid}) \quad (25)$$
$$+ (Used_{RG}(Select_{hybrid}) * Size(RowGroup))$$
$$+ (Used_{chunks}(Hybrid) * Size(Meta_{hybrid}))$$

$$Cost_{select}(Hybrid) = Used_{Chunks}(Select_{hybrid}) * W_{ReadTransfer} \quad (26)$$
$$+ Seeks(Select_{hybrid}) * (1 - W_{ReadTransfer})$$

*Hybrid layouts* They keep statistical information about data values in every column for every row group (typically, inside the header or footer sections). This helps in skipping some of the row groups that do not satisfy the predicate. Thus, the number of row groups to be read depends on the filtering condition and the sorting order of the column on which the selection is applied.

For unsorted columns, we can use the probability as in Equation 22 (borrowed from bitmap indexes [5]) to estimate the likelihood of any data in a row group satisfying the condition (i.e., a row group being fetched). In Equation 24, this probability is used to obtain the expected number of retrieved row groups. However, if a column is sorted, then we are using the *Selectivity Factor (SF)* to estimate how much data is going to be read using Equation 23, which is later used in Equation 24 to calculate the fetched row groups for sorted columns (notice that all data fulfilling the condition is stored together if they are sorted on that column). Having the number of selected row groups, Equation 25 determines the size of a selection by adding up the total size of fetched row groups, metadata, header, and footer sections. As previously discussed about multiple reads of metadata in each task, we also consider this factor in the estimation of selection size.

Finally, this selection size can be used to estimate the total number of chunks and seeks as in Equations 2 and 3, which are then weighted as in Equation 26 to estimate the total selection cost.

## 5 Experiments

In this section, we evaluate our approach and show the accuracy of our cost model for estimating the file sizes and the cost of scan, projection and selection for different data formats. We choose representative data formats from Apache Hadoop, the most popular distributed processing framework, because it is still used in 59% of the enterprises to process big data, as shown in a survey from Cloudera[2]. In order to generate realistic data-intensive workflows, we rely on standard industry benchmarks. In our previous work [20], we used TPC-H for evaluating our rule-based approach given that TPC-H provides OLAP-like queries that are typically characterized by a low selectivity factor. To properly assess our cost-model a broader range of representative analytical queries (i.e., typical query reporting and data mining are also considered) are required. For this reason, we leverage also on TPC-DS for a more representative set of experiments.

Prior to conduct our experiments we first instantiate our cost-model for Apache Hadoop. In HDFS, we can find several data formats that follow the storage layouts discussed above. Among them, we choose the most representative ones to show the effectiveness of our approach: SequenceFile (SeqFile) and Avro for horizontal layouts and Parquet for hybrid layouts. Note that, despite being included in Section 4 for the sake of completeness, we did not include any vertical layout, since those available for HDFS ended up being subsumed by hybrid ones and deprecated with time. Appendix A contains all the details about the instantiation of these formats, including the file format size calculation and the required system variables. Finally, note that for the sake of a fairer comparison, we are not considering encoding, which is available only in Parquet.

### 5.1 Setup and dataset

Our experiments are performed on a 16-machines cluster[15]. Each machine has a Xeon E5-2630L v2 @2.40GHz CPU, 128GB of main memory and 1TB SATA-3 of hard disk and runs Hadoop 2.6.2 and Pig 0.16.0 on Ubuntu 14.04 (64 bit). We have dedicated one machine for the HDFS name node and the remaining 15 machines for data nodes. We are using Apache Parquet 1.9.0, Avro 1.7.0 and elephant-bird 4.9[16] for SeqFile.

### 5.2 Validation of file size estimations

In this section, we are validating the accuracy of our size estimation by creating a materialized node (i.e., join of Lineitem and Part tables of TPC-H), and compare the actual size with the estimated one for each operation, namely scan, projection, and selection. Figure 8 shows the results for scan operation on different scale factors. Figure 8a shows the results for the size, while Figure 8b shows the corresponding error rate for each studied format. We see that Avro and Parquet are slightly underestimated (up to -3% error), but SeqFile is slightly overestimated (up to 0.5%).

---

[15] http://www.ac.upc.edu/serveis-tic/altas-prestaciones
[16] https://github.com/twitter/elephant-bird

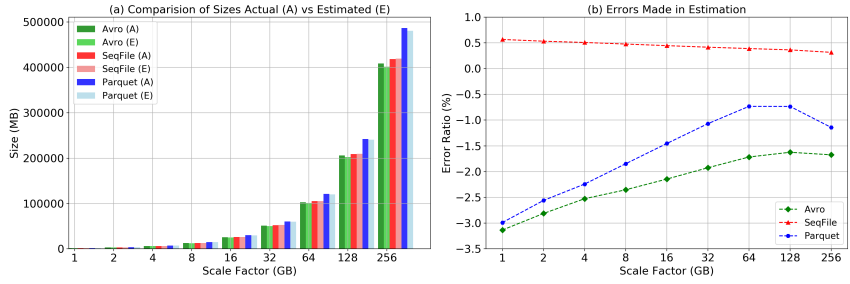

**Fig. 8** Validating the size estimation

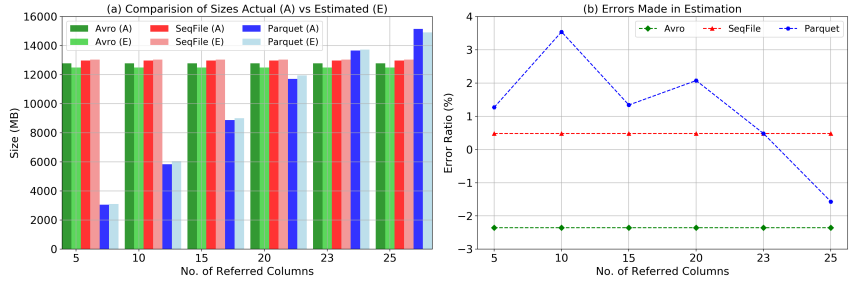

**Fig. 9** Validating the projection cost model

Similarly, Figure 9 shows the results for validating our file size estimation after a projection. To do so, we read different number of columns, ranging from 5 to 25, by executing 100 different runs, selecting different columns on each run, over 8GB and took the average of all runs. Figure 9a compares the actual and estimated size, and Figure 9b shows the percentage of error. SeqFile and Avro do the scan for projection and their errors are the same as of the scan. However, Parquet has errors between +4% to -2%, whose variance is due to variable column sizes (e.g., column with string data type), whereas we use average column size for all columns.

Finally, Figure 10 validates the file size after a selection operation. For this experiment, we generate different selectivity factors. Also, since, the sorting order of the filter column affects the reading, we are validating our results for both sorted and unsorted columns. Moreover, we repeated our experiments 100 times over 8GB by randomly choosing different search values and took the average of all executions. Figures 10a and 10b show the results for unsorted columns, which show that our cost model underestimates the sizes (i.e., upto -4%). The reason for that underestimation in the presence of small selectivity factors is due to the distribution of data among row groups, which is less variable the more data we retrieve. Thus, with few data, the variation is high, and consequently harder to predict. On the other hand, Figures 10c and 10d show the results for sorted columns. Here our cost model constantly underestimates Parquet in the range of +2% to -4% for the same reason discussed for unsorted columns.

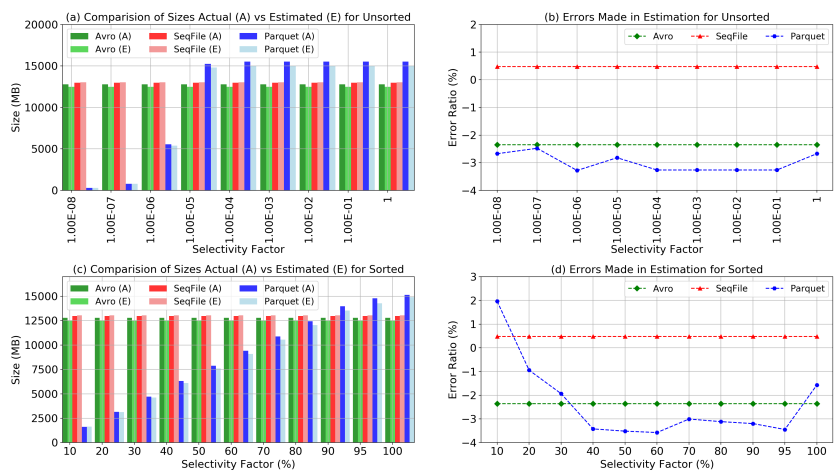

**Fig. 10** Validating the selection cost model

All in all, the errors obtained in all our tests are rather small and consistent. Most importantly, we show next that these errors do not affect our prediction to choose the right storage format in all the experiments we conducted, since the estimated values always preserve the partial order among the actual values.

5.3 Validation of file format choice

In our previous work [20], we utilized TPC-H (i.e., OLAP-like workloads) for validating the accuracy of our heuristic rules and observed the importance of workload and data characteristics in selecting the most appropriate format. In this paper we propose a cost model and validate it with both TPC-H and TPC-DS benchmarks. The goal is to cover a broader range of queries (i.e., broader workloads spanning reporting, OLAP and data mining).

In order to create a complex DIW, we used Quarry [18] to combine all TPC-DS queries into one integrated DIW as shown in Figure 11. To perform realistic experiments, we generate data with scale factors ranging from 1GB to 256GB. In our experiments, ReStore (see Section 2.2) is used and nine nodes are selected to be materialized after applying both its aggressive and conser-vative heuristics. The aggressive heuristics decide to materialize the output of six joins and the conservative heuristics add three nodes resulting of filter operations. Additionally, we choose two metrics to analyze our approach, namely write cost (Section 4.1) and read cost (Section 4.2) for each materialized node. However, due to limitations in the native measurement of Hadoop performance, the charts corresponding to the read cost include also the execution cost of the first operation right after reading the IR, since their costs cannot be decoupled.

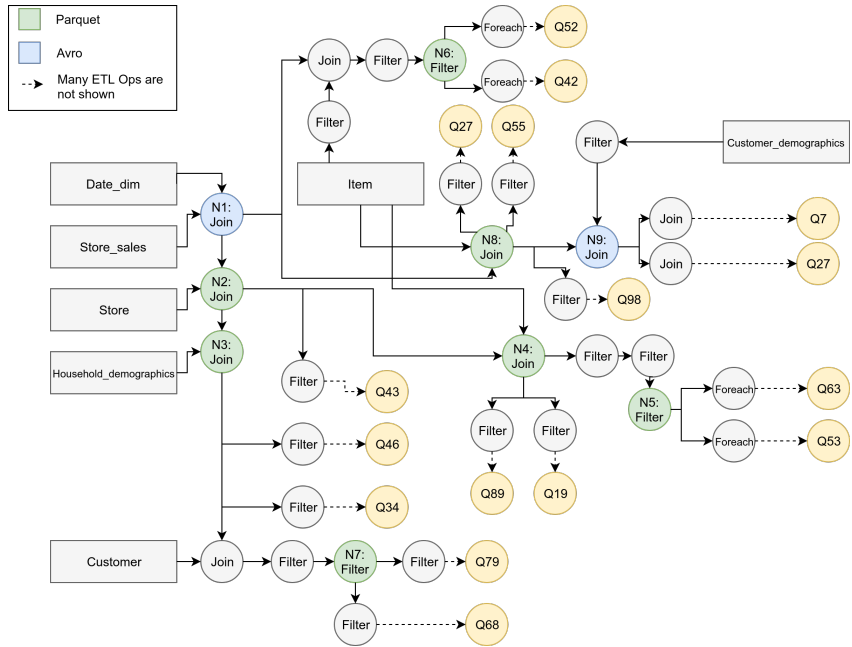

**Fig. 11** DIW of 16 TPC-DS queries

*Rule-based approach* Table 2 shows all nine nodes that have been materialized, together with their outgoing operators and storage formats decided by applying the heuristic rules from [20]. Avro is chosen for N1 and N9, because the outgoing operators are joins, that use a scan access pattern, where Avro excels, as discussed in Section 2.1. For all other nodes, the rule-based approach is choosing Parquet. For Nodes N5 and N6, the outgoing edges contain FOREACH operations, where Parquet benefits from independent column storage. Nodes N4, N7 and N8 have FILTER operations in their outgoing edges, where Parquet can benefit from its native predicate push-down. Both FOREACH and FILTER operations only require a subset of data, and Parquet excels whenever a subset of data is read. Finally, nodes N2 and N3 have JOIN and FILTER as outgoing edges, and there would be different options to choose.

Table 2 Materialized nodes with the statistics about their operations and chosen storage formats

| Node | Outgoing Operators | Rule-based | Cost-based | Real Best Choice |
|---|---|---|---|---|
| N1 | JOIN, JOIN | Avro | Avro | Avro |
| N2 | JOIN, JOIN, FILTER (SF: 0.19) | Parquet | Avro | Avro |
| N3 | JOIN, FILTER (SF: 0.59), FILTER (SF: 0.01) | Parquet | Avro | Avro |
| N4 | FILTER (SF: 0.03), FILTER (SF: 0.2), FILTER (SF: 0.19) | Parquet | Avro | Avro |
| N5 | FOREACH (Ref Cols: 3), FOREACH (Ref Cols: 3) | Parquet | Parquet | Parquet |
| N6 | FOREACH (Ref Cols: 4), FOREACH (Ref Cols: 4) | Parquet | Parquet | Parquet |
| N7 | FILTER (SF: 0.13), FILTER (SF: 0.92) | Parquet | Avro | Avro |
| N8 | JOIN, FILTER (SF: 0.19), FILTER (SF: 0.03), FILTER (SF: 0.01) | Parquet | Avro | Avro |
| N9 | JOIN, JOIN | Avro | Avro | Avro |

*Projection is implemented as FOREACH in Apache PIG

However, Parquet is preferred in front of Avro, since, in case of several options available, our rule-based approach chooses the richest format providing more features.

*Cost-based approach* Note that Table 2 also shows some relevant collected statistics, such as the selectivity factor (SF) and the number of referred columns (Ref Cols), of the outgoing operators. Moreover, we have divided these nine nodes into three different color groups which are green, grey, and white. Green and grey groups contain nodes for which our rule-based approach works fine. Whereas, white group contains all the nodes for which our rule-based approach fails.

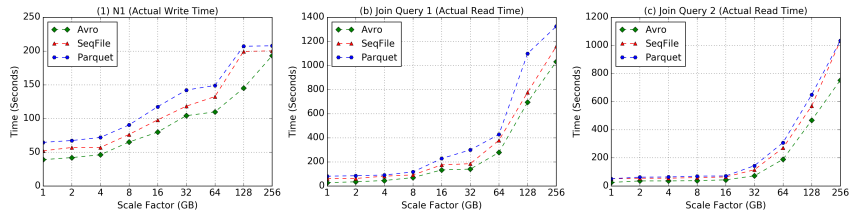

Fig. 12 Detailed experimentation conducted for node N1

Let us focus on N1 from green group, for which the rule-based approach chooses the correct storage format (i.e., Avro). Figure 12 scrutinizes the actual write / read time of each storage format. It can be verified that the chosen layout (i.e., Avro) is always faster for both write and read operations.

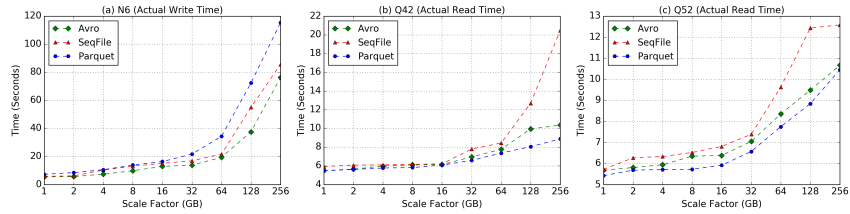

**Fig. 13** Results for N6

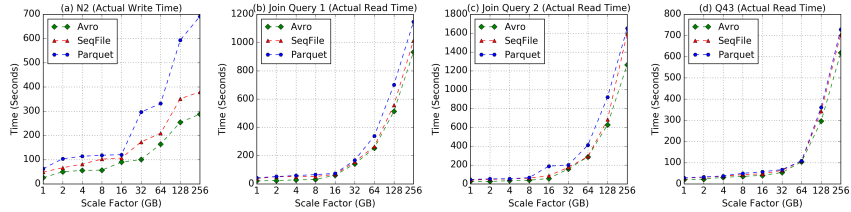

**Fig. 14** Results for N2

Similarly, rule-based approach also chooses the right storage format for grey group. Grey group contains nodes with projection operation. However, the amount of data read is less than 70% and that's why it is better to use Parquet. Figure 13 shows the results for N6 and it can be seen that our cost-based approach also chooses Parquet. Figure 13a shows the actual execution time for both write and read operations. It can be seen that Parquet takes more time in writing (i.e., it writes more metadata), but its benefits compensate in reading as shown in Figures 13b and 13c.

On the other hand, the rule-based approach failed to choose the correct storage formats for white group. All these nodes involve filter operations, where the amount of data to be read depends on the selectivity factor and all of them are greater than or equal to 0.1 (see Table 2). As already shown in Figure 10, different storage formats perform differently depending on the amount of data read. Therefore, since the rule-based approach does not leverage on statistics, the data volume to be read is not considered and it fails when choosing the right storage format. As discussed in Section 5.2, the predicate push-down mechanism implemented by Parquet is useless when the selectivity factor is greater than *1.0E-05* for unsorted columns. However, the rule-based approach always considers predicate push-down to be worth and thus still chooses Parquet. Oppositely, since our cost-based model considers the selectivity factor, it is able to select the right format for these nodes. For example, the results for N2 of white group are shown in Figure 14, where the optimal choice is Avro, which takes less time than Parquet in both write and read operations. The prediction of our cost model can be verified by the actual execution which is shown in Figure 14. All the nodes of white group follow the same trend and our cost-based approach successfully choses the right storage format.

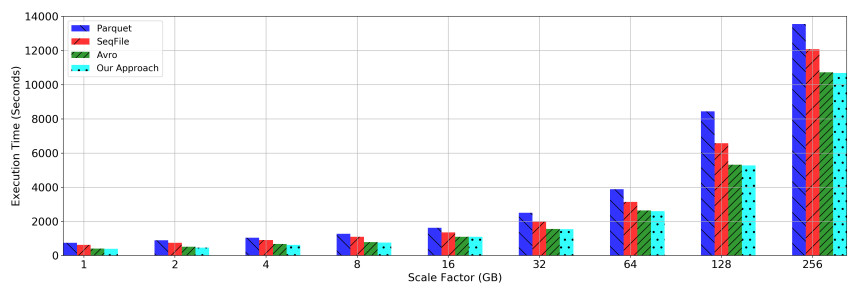

**Fig. 15** Single Fixed Format vs Our Approach for TPC-DS

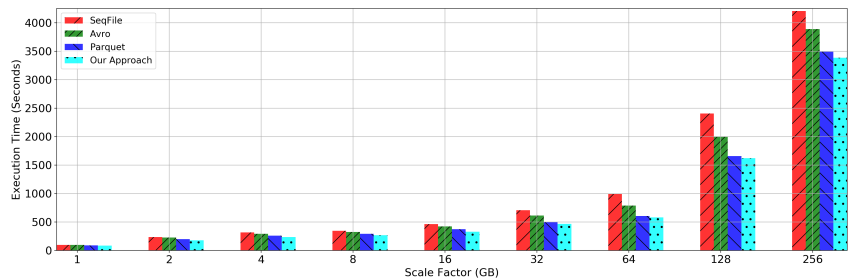

**Fig. 16** Single Fixed Format vs Our Approach for TPC-H

In general, our cost model is able to decide the right format in all cases shown in Table 2, because it considers the amount of data read (which in this case is determined by the format file size and the selectivity factor of the operation), which actually depends on two operations, namely projection and selection. Figure 15 compares our approach with a typical approach materializing all chosen IR with a fixed format (i.e., always SequenceFile, Parquet or Avro). It shows the overall execution time of the DIW when using a single fixed format for materialized IR with regard to a dynamic choice of the format based on our cost model. Our approach on average provides 60% speed up over fixed Parquet, 34% speedup over fixed SequenceFile, 3% speedup over fixed Avro and, in the average, it provides 33% speedup for TPC-DS.

Note that in TPC-DS, our cost model favors Avro, and this is due to the fact that the chosen materialized nodes have subsequent operations with high selectivity factors. In contrast, when we changed workload to TPC-H, the cost model recommends Parquet in majority of the materialized nodes, due to the low selectivity queries. The overall results of TPC-H are shown in Figure 16. Observe that, for TPC-H, our approach on average provides 32% speedup over fixed SequenceFile, 19% speedup over fixed Avro, 4% speedup over fixed Parquet and overall, it provides 18% speedup.

In conclusion, for different workloads our cost model capable of choosing the appropriate storage format, which is always lead to improvements in query execution time.

# 6 Related Work

In this section, we first discuss the related work on choosing different storage layouts. Then, we discuss in detail the existing cost models for distributed processing frameworks.

## 6.1 Use of different storage layouts

The fixed layout problem is already identified by the research community and many solutions have been proposed. The existing solutions allow using multiple layouts together. For instance, the in-memory DBMS SAP HANA [10] uses horizontal and vertical layouts for On-line Transaction Processing (OLTP) and On-line Analytical Processing (OLAP) workloads, respectively. In a similar way, in DB2 [22] horizontal and vertical layouts can be used for the same table-space. However, these layouts are fixed and non-modifiable at runtime.

On the other hand, there are also solutions that consider workloads in order to decide for the most suitable layout. These systems work in multi-database environments. Polybase [7] for instance, is a system that uses both Hadoop cluster and DBMS for data storage. Based on the workloads, it dynamically decides which is the best solution. According to this decision, it also moves the data from one system to another for the execution of the queries. This solution focuses on utilizing the processing power of the Hadoop cluster and it always uses a horizontal layout to store data on Hadoop.

Similar to Polybase, there is a hybrid system [13], which can read raw files directly and it can choose the physical layout of the data in the DBMS based on the input queries. However, they propose to keep multiple copies of the same data in different formats, which is not always feasible, especially when the size of the data is huge.

In addition, there are two systems [9,23] that store the data inside different storage engines by taking into account the data access patterns. These systems work like mediators and analyze the characteristics of the data to then route them to the most suitable storage engine. In [9], the system requires training in order to take the right decision in choosing the best storage engine for queries. Furthermore, this training runs every query in all available systems to see which system is good for most of the queries. Hence, this requires extra processing and adds extra cost. In [23], the solution relies on annotations which are defined by the user during the requirements definition process of an application. These annotations help the mediator to decide where to store the application data. The annotations however cannot be defined at run-time. Moreover, this solution mainly focuses on choosing a storage engine according to the application requirements without considering the physical storage layout.

There is a solution, H2O [1], that can dynamically decide the physical layout of the data based on the current workload. However, it only supports vertical layouts by creating different column groups. Moreover, as described there, creating column groups is a NP-hard problem and it is not feasible for a table which has many columns. Additionally, WWHow [17] proposes a data layer which is independent of the physical storage. This layer enables an adaptable physical storage engine by analyzing the application needs. However, they are considering general storage systems such as file-systems, databases and cloud storages without considering the physical layouts of the data in them. Moreover, once decided, the storage system remains fixed. [3] proposes a caching approach for nested data (i.e., JSON). It helps to keep more frequent used data in cache by storing in appropriate layout, according to the running workload. This work also supports our hypothesis to use different layouts for different type of workloads. However, it is limited to only nested data and not applicable to other scenarios.

6.2 Existing cost models

Trojan [16] is an adaptable column storage for Hadoop that handles different types of workloads. It takes advantage of the data replication feature of HDFS, and analyzes the workload access patterns to store different column groups on each replica. Then, it routes every query to the node where the replica of the data has the most suitable layout for that particular query. However, it considers only the vertical storage layouts and ignores scan based workloads. They also proposed a cost model for different storage layouts, but their cost model considers only the scan operation.

Furthermore, there is also a cost model for Hadoop jobs, Starfish [12], whose cost model helps to measure execution time. It considers different parameters to calculate the execution time, and it can help to design a cost-based optimizer. However, it does not consider different storage layouts.

Finally, [4] helps in reducing the seek cost in a wide table by storing the columns in an appropriate order based on the access patterns. This approach helps to reduce the disk cost and, overall, it reduces the execution cost of different queries. However, it considers only hybrid layouts in their study and it provides a cost model only for estimating the seek cost.

7 Conclusion

Modern analytical workloads involve different types of queries in which a fixed data format for materializing output of common tasks does not guarantee the best performance. We propose an approach that helps choosing the best data format based on the features of the subsequent operations consuming such materialized output. Accordingly, after deciding which nodes in a data intensive workflow to materialize, we choose the best storage format, which improves performance, by analyzing their access patterns. Overall, this reduces the load time and, in general, the total workflow execution time. We have implemented our generic cost-based model for Hadoop to show its effectiveness. Our evaluation results show the benefits of our approach and support our hypothesis that intermediate results should be materialized by considering the best storage format for each of them.

**Acknowledgements** This research has been funded by the European Commission through the Erasmus Mundus Joint Doctorate "Information Technologies for Business Intelligence - Doctoral College" (IT4BI-DC)

# A

This appendix shows the file sizes for the three considered HDFS file formats, together with the system variables with their values according to our testbed. Table 3 lists all the system variables. They are divided in three categories. First category has the variables related to disk which are important to calculate the reading and writing cost. Additionally, second category has variables for network to calculate the transfer cost, since Hadoop writes multiple copy of data for fault tolerance purpose and this involves writing to other nodes. For this writing, it needs to transfer data, and it is important in calculating the overall write cost. Final category lists the variables related to the configuration of our Hadoop cluster.

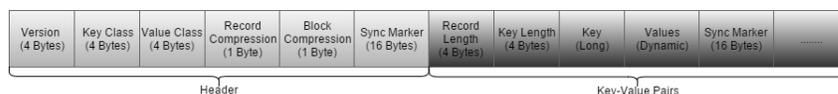

**Fig. 17** Physical file format of SequenceFile

## A.1 SequenceFile (SeqFile) format

SeqFile[17] is introduced in 2009 to improve the performance of MapReduce framework. It is used to store the temporary output of map phases as compressed to reduce I/Os. Moreover, it is also splittable which is ideal for processing in parallel. It considers a special type of horizontal layout, which stores data in the form of key-value pairs. Figure 17 shows its structure and Table 4 shows the specific variables of SeqFile with their values.

---

[17] https://wiki.apache.org/hadoop/SequenceFile

**Table 3** System variables with their values according to our testbed

| | Variables for Disk | |
|---|---|---|
| $BW_{disk}$ | Disk bandwidth | $1.3 \times 10^8$ bytes/second |
| $Size(Block)$ | Disk block size | $8.0 \times 10^3$ bytes |
| $Time_{seek}$ | Disk random seek time | $5.0 \times 10^{-3}$ seconds |
| $Time_{rotation}$ | Disk rotation time | $4.17 \times 10^{-6}$ seconds |
| | Variables for Network | |
| $BW_{net}$ | Network bandwidth | $1.25 \times 10^8$ bytes/second |
| | Variables for Hadoop | |
| $Size(Chunk)$ | HDFS block size | $1.28 \times 10^8$ bytes |
| $Size(Buffer)$ | Buffer size | $6.4 \times 10^4$ bytes |
| $R$ | Replication factor | 3 |
| $p$ | Probability of accessed replica being local [16] | 0.97 |

**Table 4** Sizes of SeqFile according to our testbed

| | Variables for SeqFile | |
|---|---|---|
| $Size(Header_{SeqFile})$ | Header size of SeqFile | 30 |
| $Size(RecordLength)$ | Fixed field | 4 |
| $Size(KeyLength)$ | Number of bytes for key | 4 |
| $Size(Meta_{SCol})$ | Number of bytes for user-defined separator per column | 1 |
| $Size(SyncMarker)$ | Size of sync marker | 16 |
| $Size(SyncBlock)$ | Number of bytes between sync markers | 2,000 |
| $Size(Footer_{SeqFile})$ | Footer size | 0 |

$$Size(Row_{SeqFile}) = Size(RecordLength) \\ + Size(KeyLength) \\ + Size(Col) * Cols(IR) \\ + Size(Meta_{SCol}) * (Cols(IR) - 2) \qquad (27)$$

$$Size(TotalRows_{SeqFile}) = Size(Row_{SeqFile}) * |IR| \qquad (28)$$

$$Size(Meta_{SBody}) = \left\lceil \frac{Size(TotalRows_{SeqFile})}{Size(SyncBlock)} \right\rceil \\ * Size(SyncMarker) \qquad (29)$$

$$Size(Body_{SeqFile}) = Size(TotalRows_{SeqFile}) \\ + Size(Meta_{SBody}) \qquad (30)$$

To instantiate from our generic cost model, we need to estimate the sizes of header, body and footer sections. The header section of SeqFile has a fixed size, so we define it as a constant. To estimate body size, we need to calculate row and metadata sizes. SeqFile divides each row into a key-value pair and stores one column into the key, and the remaining columns into the value by using a user-defined separator. Thus, it has two types of metadata: one is used to separate values and another to make blocks for parallel processing. Then, the size of a row is compound of some fields of fixed size (i.e., record and key lengths) together with the corresponding key-value pair as shown in Figure 17, containing all user columns (notice that we need two less user-defined separators than columns, because the key is managed by the file format itself). Equation 27 is estimating this size (i.e., a row for SeqFile), which is later used in Equation 28 to estimate the size of all key-value pairs. Equation 29 calculates the overhead of block-related metadata (i.e., sync markers), which SeqFile introduces at fixed intervals. Finally, Equation 30 simply adds the size of key-value pairs and metadata, which allows in turn to obtain the total size of SeqFile using Equation 1 with an empty footer section.

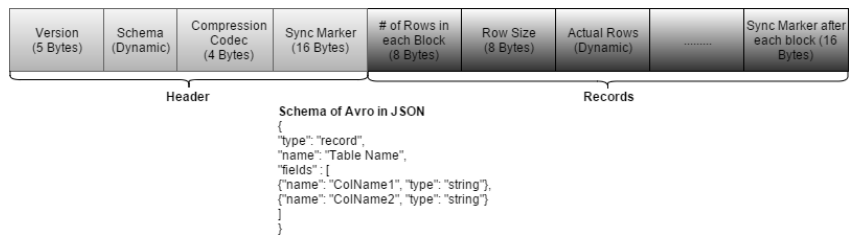

**Fig. 18** Physical file format of Avro

## A.2 Avro format

**Table 5** Sizes of Avro according to our testbed

| | Variables for Avro | |
|---|---|---|
| $Size(Version)$ | Version of Avro | 5 |
| $Size(Codec)$ | Compression codec | 4 |
| $Size(SyncMarker)$ | Size of sync marker | 16 |
| $Size(ColSchema)$ | Size of schema information per column | ~30 bytes |
| $Size(B_{avro})$ | Block size of Avro | 4,000 |
| $Size(Meta_{ARow})$ | Meta information for each row | 8 |
| $Size(Meta_{ABlock})$ | Meta information for each block | 8 |
| $Size(Footer_{Avro})$ | Footer size | 0 |

Apache Avro[18] is a language-neutral data serialization system. It means Avro can be written in one language and can be read in another language without changing the code. This support is provided by the schema information which Avro stores as a meta information.

---

[18]https://www.tutorialspoint.com/avro/avro_tutorial.pdf

Moreover, it is also compressible and splitable. It is a horizontal layout and Figure 18 sketches its physical structure. Moreover, there are specific variables for Avro which are given in Table 5. The data schema is stored in a header section of variable length. Similarly, the size of body is also variable and it depends on the number of rows in an IR.

$$Size(Header_{avro}) \quad = Size(Version) \qquad (31)$$
$$+ Cols(IR) * Size(ColSchema)$$
$$+ Size(Codec) + Size(SyncMarker)$$

$$Size(TotalRows_{Avro}) = (Size(Row) + Size(Meta_{ARow})) * |IR| \qquad (32)$$

$$Size(Meta_{ABody}) \quad = (Meta_{ABlock} + Size(SyncMarker)) \qquad (33)$$
$$* \left\lceil \frac{Size(TotalRows_{Avro})}{Size(B_{Avro})} \right\rceil$$

$$Size(Body_{avro}) \quad = Size(TotalRows) + Size(Meta_{ABody}) \qquad (34)$$

Header section of Avro contains meta information corresponding to the schema of the data in the form a JSON. Given that the size of the schema is orders of magnitude smaller that data, we estimate it as a constant per column. Considering also the version and codec information, the overall header size is calculated by Equation 31. Following the horizontal layout, Avro adds metadata to each row, which is considered in Equation 32 to estimate the size of a row. Moreover, it also adds extra metadata in the body for every block. Thus, Equation 33 is calculating the total size of metadata by multiplying the number of blocks by the size of sync marker and that of counter for the number of rows in the block. Finally, Equation 34 is used to calculate the body size, which allows in turn to obtain the total size of Avro using Equation 1 with an empty footer section.

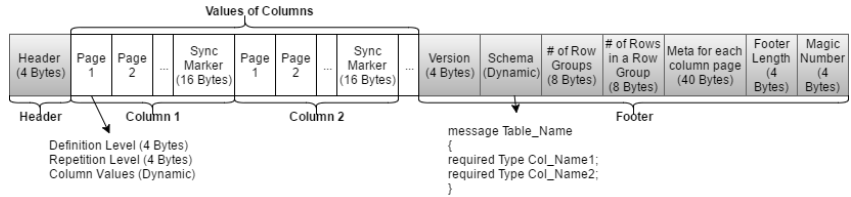

**Fig. 19** Physical file format of Parquet

### A.3 Parquet format

Apache Parquet[19] is introduced in 2013 to provide hybrid layout support for Hadoop echosystem. It divides data horizontally into row groups, whereas each row group is further divided vertically to store columns separately, as sketched in Figure 19. Additionally, it also divides each vertical partition into multiple pages. Moreover, it also stores the schema and statistical information about the data as meta information in the footer section. All variables specific to Parquet are listed in Table 6.

---

[19] http://parquet.apache.org

Table 6 Sizes of Parquet according to our testbed

| Variables for Parquet | | |
|---|---|---|
| $Size(Header_{parquet})$ | Size of header | 4 |
| $Size(DefinitionLevel)$ | Size of definition level | 4 |
| $Size(RepetitionLevel)$ | Size of repetition level | 4 |
| $Size(RowCounter)$ | Size of number of rows | 8 |
| $Size(SyncMarker)$ | Size of sync marker | 16 |
| $Size(Version)$ | Version in footer | 4 |
| $Size(ColSchema)$ | Size of schema information per column | ~30 bytes |
| $Size(Meta_{PCol})$ | Size of columns meta data for storing statistical information | 40 |
| $Size(MagicNumber)$ | Magic number in footer | 4 |
| $Size(FooterLength)$ | Footer length in footer | 4 |
| $Size(RowGroup)$ | Layout row group size | $1.28 \times 10^8$ bytes |
| $Size(Page)$ | Layout page size | $1.05 \times 10^6$ bytes |

$$Used_{pages}(RowGroup_{parquet}) = (Size(Col) \qquad (35)$$
$$* Used_{rows}(RowGroup_{parquet})$$
$$+ Size(SyncMarker))$$
$$* \frac{Cols(IR)}{Size(Page)}$$

$$Size(Body_{parquet}) = (((Size(DefinitionLevel) \qquad (36)$$
$$+ Size(RepetitionLevel)$$
$$+ Size(Page))$$
$$* Used_{pages}(RowGroup_{parquet}))$$
$$+ Size(RowCounter)$$
$$+ Size(SyncMarker))$$
$$* Used_{RG}(Parquet)$$

$$Size(Footer_{parquet}) = Size(Version) \qquad (37)$$
$$+ Size(ColSchema) * Cols(IR)$$
$$+ Size(MagicNumber)$$
$$+ Size(FooterLength)$$
$$+ Used_{RG}(Parquet) * Size(Meta_{PCol})$$
$$* (1 + Used_{pages}(RowGroup_{parquet}))$$

The header section of Parquet has a fixed size, as stated in Table 6. To estimate the body size, we first need to estimate the total number of row groups (i.e., Equation 9) and the total rows per row group (i.e., Equation 18). Moreover, we need to be aware that Parquet stores every individual column divided it into multiple pages, whose number which is estimated by Equation 35 per row group. Next, we are calculating the body size of Parquet using Equation 36, by considering metadata for each page (namely definition level and repetition level), and for every row group (namely counter of rows per row group and sync marker).

Finally, we calculate the footer size by approximating the size the of the schema, sketched in Figure 19, by a constant amount of bytes per column. Moreover, Parquet also stores statistical information about columns in the Footer section for both row groups and data pages. Equation 37 uses all these values together to calculate overall size of footer. Then, total size of Parquet is obtained by adding the header, body and footer sections, as defined in Equation 1.